\documentstyle[12pt]{article}
\begin{document}
\baselineskip = 24pt

\begin{titlepage}
\vspace{4.5cm}
\begin{flushright} ITP-SB-93-66 \\
\end{flushright}

\begin{center}{\large \vspace{1.0cm} \bf The Yangian symmetry of the Hubbard
Model}\\
\vspace{1cm}
D.B.Uglov \footnotemark[1] and V.E.Korepin \footnotemark[2] \\
\footnotetext[1]{e-mail: denis@max.physics.sunysb.edu}
\footnotetext[2]{e-mail: korepin@max.physics.sunysb.edu}
\vspace{1cm}
Institute for Theoretical Physics, State University of New York at Stony Brook
\\
Stony Brook, NY 11794-3840, USA \\
\vspace{0.5cm}
October 23, 1993
\vspace{3cm}
\begin{abstract}
We discovered new hidden symmetry of the one-dimensional Hubbard model. We show
that the one-dimensional Hubbard model on the infinite chain has the
infinite-dimensional algebra of symmetries. This algebra is a direct sum of two
$ sl(2) $-Yangians. This $ Y(sl(2)) \oplus Y(sl(2)) $ symmetry is an extension
of the well-known $ sl(2) \oplus sl(2) $ . The deformation parameters of the
Yangians are equal up to the signs  to the coupling constant of the Hubbard
model hamiltonian. \end{abstract}

\end{center}

\vspace{3cm}

\end{titlepage}

\section{Introduction}
Models of strongly correlated electronic systems are currently under intense
study in relation to high-$ T_{c} $ superconductivity. The Hubbard model is in
the very center of these investigations. A number of results pertaining to it
are now available. For a guidance in the extensive literature devoted to the
Hubbard model we  refer a reader to several reviews and reprint volumes [3] and
the references therein.

The first exact results in the one-dimensional Hubbard model were obtained by
E. Lieb and F. Wu [1], who were using the Nested Bethe-Ansatz technique
discovered in [2] . Since then, through the work of many authors, this model
acquired a status of exactly solvable one.  All of the progress achieved so far
in the exactly solvable one-dimensional Hubbard model came through the use of
the coordinate Bethe-Ansatz in combination with the $ sl(2) \oplus sl(2) $
internal symmetry [12] exhibited by the Hubbard model hamiltonian. In
particular in this way the exact S-matrix of the model was recently found [4].

Though the long-distance asymptotics of the correlaton functions in the Hubbard
model are very well studied[5], exact formulas for them have, so far, been
unaccessible by means of the traditional Bethe-Ansatz-based approach. In this
light, in order to get the exact correlators and form-factors, it seems natural
to turn to the symmetry-based approach which has been recently pioneered with
impressive success in the series of works [6-8] dealing with the wide range of
models, most notably XXZ magnet. In this approach one starts with a given model
directly on infinite lattice and looks for an ifinite-dimensional algebra of
symmetries ( infinite dimensional Lie algebra or some quantum group which is a
deformation of an infinite dimensional Lie algebra) which, if large enough,
permits to solve the model completely in the spirit close to that one of
Conformal Field Theory. To carry out such a program for the Hubbard model one
first has to identify its algebra of symmetries on the infinite chai
   n.  One comes up with
the result that this algebra contains a pair of commuting $sl(2)$-Yangians [9].
Each of these Yangians is an infinite dimensional quantum group which is a
deformation of the non-negative frequency part of the $ sl(2) $ Kac-Moody
algebra, absolute values of the deformation parameters and the coupling
constant of the Hubbard model being equal. This Yangian symmetry, probably,
admits an extension to the Yangian Double [10], however we don't consider this
question in the present letter. Important motivation to look for the Yangian
symmetry in the Hubbard model came from the paper [4], where the two-particle
S-matrix of the elementary excitations was found. This S-matrix is easily
recognizable as an intertwiner between the tensor products of the two
(simplest) evaluation representations [11] of the direct sum of two Yangians.
This pointed out the likely existence of the Yangian symmetry in the model (
For the detailed general discussion of this point see, for example, [10] ).

\section{The Yangian symmetry of the Hubbard model}
In this section we give explicit expressions for the generators of the two
Yangians commuting with the Hubbard hamiltonian on the infinite chain. We omit
quite long and tedious ( but straitforward ) computations leading to the result
formulated below.

The Hubbard hamiltonian we are working with is given by the following
expression:
\begin{equation}
H = -\sum_{i} \sum_{\sigma = \uparrow,\downarrow}(c_{i,\sigma}^{\dagger}
c_{i+1,\sigma} + c_{i+1,\sigma}^{\dagger} c_{i,\sigma}) +
U\sum_{i}(n_{i,\uparrow}-\frac{1}{2})(n_{i,\downarrow}-\frac{1}{2}) .
\end{equation}
Here $ c_{i,\sigma} $ are canonical fermionic annihilation operators on the
lattice, $i$ runs through the integer numbers and labels the sites of the
infinite chain, $ \sigma $ labels the spin degrees of freedom, $ U $ is the
(real) coupling constant, and $ n_{i,\sigma}=c_{i,\sigma}^{\dagger}c_{i,\sigma}
$ is the number operator for spin $ \sigma $ on site $ i $ .

In order to formulate the result we introduce the following notations for the
densities of the symmetry generators:
\begin{equation}
{\cal E}_{i}^{n}= c_{i,\uparrow}^{\dagger}c_{i+n,\downarrow} \;, \; \;{\cal
F}_{i}^{n}= c_{i,\downarrow}^{\dagger}c_{i+n,\uparrow} \;, \; \;{\cal
H}_{i}^{n}= c_{i,\uparrow}^{\dagger}c_{i+n,\uparrow} -
c_{i,\downarrow}^{\dagger}c_{i+n,\downarrow}.
\end{equation}

Now we formulate the principal:

\newtheorem{stat}{Statement}

\begin{stat}
{\sl the six operators: }$ (E_{0}, F_{0}, H_{0}, E_{1}, F_{1}, H_{1}) $ {\sl
given below commute with the hamiltonian (1) }:
\begin{equation}
E_{0} = \sum_{i} {\cal E}_{i}^{0} \;, \; \; F_{0}= \sum_{i} {\cal F}_{i}^{0}\;,
\; \; H_{0}=\sum_{i} {\cal H}_{i}^{0} \;,
\end{equation}
\begin{eqnarray}
E_{1}& =& \sum_{i} ( {\cal E}_{i}^{1} - {\cal E}_{i}^{-1} ) -
\frac{U}{2}\sum_{i<j}({\cal E}_{i}^{0}{\cal H}_{j}^{0} -   {\cal
E}_{j}^{0}{\cal H}_{i}^{0} ) \; ,  \\
F_{1}& = &\sum_{i} ( {\cal F}_{i}^{1} - {\cal F}_{i}^{-1} ) +
\frac{U}{2}\sum_{i<j}({\cal F}_{i}^{0}{\cal H}_{j}^{0} -   {\cal
F}_{j}^{0}{\cal H}_{i}^{0} ) \; ,  \\
H_{1}& =& \sum_{i} ( {\cal H}_{i}^{1} - {\cal H}_{i}^{-1} ) + U\sum_{i<j}({\cal
E}_{i}^{0}{\cal F}_{j}^{0} -   {\cal E}_{j}^{0}{\cal F}_{i}^{0} ) \; .
\end{eqnarray}
\end{stat}

Verification of this statement is a straightforward calculation.
The above operators, as we will show in a moment, give us the generators of one
of the Yangians. In order to get another one we proceed as follows: one easily
verifies that under the canonical transformation :
\begin{equation}
 c_{i,\downarrow} \rightarrow c_{i,\downarrow}\;, \; \; c_{i,\uparrow}
\rightarrow (-1)^{i}c_{i,\uparrow}^{\dagger}
\end{equation}
the kinetic term in (1) doesn't change, while the interaction term changes its
sign. This observation immediately leads to the conclusion that the operators $
E_{0}^{'}, F_{0}^{'}, H_{0}^{'}, E_{1}^{'}, F_{1}^{'},H_{1}^{'} $ obtained from
(3-6) under the transformation (7) and $ U \rightarrow -U $ commute with the
hamiltonian as well. Thus we obtained 12 generators commuting with the Hubbard
hamiltonian (1).

Now we clarify what algebra these generators generate. First of all the
operators (3) and their primed counterparts are the well-known $ sl(2) \oplus
sl(2) $ generators. They are known to commute with the Hubbard hamiltonian
(under the variety of boundary conditions) on the finite chain also,  as well
as in higher dimensions [12]. The rest of the generators extend $ sl(2) \oplus
sl(2) $ to $ Y(sl(2)) \oplus Y(sl(2)) $ , so that all the unprimed generators
commute with the primed ones, and between themselves the (un)primed generators
satisfy the defining relations of the $sl(2)$-Yangian [9,10]. We formulate this
as :
\begin{stat}
i). The operators $\{ E_{\epsilon},F_{\epsilon}, H_{\epsilon} \}_{\epsilon=0,1}
$ provide a representation of the $ sl(2) $-Yangian with the deformation
parameter equal to $ U $ ( or $ -U $ ), i. e. the following defining relations
[9,10] are satisfied:
\begin{equation}
[E_{0},F_{0}]=H_{0}\;,\;\; [H_{0},E_{0}]=2E_{0}\;,\;\; [H_{0},F_{0}]=-2F_{0}.
\end{equation} ( i.e. $ E_{0}, F_{0}, H_{0} $ generate $ sl(2) $ );
\begin{equation}
[E_{0},F_{1}] = H_{1}\;,\; \; [F_{0},E_{1}]  =  -H_{1}\;,\; \;  [H_{0},E_{1}]
=  2E_{1}\;, \end{equation}
\begin{equation}
[E_{0},H_{1}]  =  -2E_{1}\;,\;\;  [F_{0},H_{1}]  = 2F_{1}\;, \;\;
[H_{0},F_{1}]  =  -2F_{1}\;, \end{equation}
\begin{equation} \\ \,
[E_{0},E_{1}]  =  0 \;,\;\;   [F_{0},F_{1}]  =  0 \;,\;\;  [H_{0},H_{1}]  =  0
\;. \end{equation}( $ E_{1},F_{1},H_{1} $ form a vector representation of (8)
);

and finally one has the deformed Serre relations:
\begin{eqnarray} \,[H_{1},[E_{1},F_{1}]] & = & U^{2} ( \{ H_{0},E_{0},F_{1} \}
- \{ H_{0},F_{0},E_{1} \} ) ,  \\  \,[E_{1},[H_{1},E_{1}]] & = & U^{2} ( \{
E_{0},H_{0},E_{1} \} - \{ E_{0},E_{0},H_{1} \} ) ,  \\  \,
[F_{1},[H_{1},F_{1}]] & = & U^{2} ( \{ F_{0},H_{0},F_{1} \} - \{
F_{0},F_{0},H_{1} \} ) ,
\end{eqnarray}
\begin{eqnarray}[H_{1},[H_{1},E_{1}]] + 2[E_{1},[E_{1},F_{1}]] = \hspace{7cm}
\nonumber \\   U^{2} (
\{H_{0},H_{0},E_{1}\}-\{H_{0},E_{0},H_{1}\}+2\{E_{0},E_{0},F_{1}\}-2\{E_{0},F_{0},E_{1}\} ) , \\ \,[H_{1},[H_{1},F_{1}]] + 2[F_{1},[F_{1},E_{1}]]  = \hspace{7cm} \nonumber \\  U^{2} ( \{H_{0},H_{0},F_{1}\}-\{H_{0},F_{0},H_{1}\}+2\{F_{0},F_{0},E_{1}\}-2\{F_{0},E_{0},F_{1}\} ) .
\end{eqnarray}
Where the curly brackets mean symmetrization: $ \{ x_{1},x_{2},x_{3} \} =
\frac{1}{6} \sum_{i_{1} \neq i_{2} \neq i_{3} } x_{i_{1}}x_{i_{2}}x_{i_{3}} $.

ii). The primed generators satisfy the same relations as above and commute with
the unprimed ones.
\end{stat}

Now let us make some comments about the results. As one sees from the above
defining  relations, one can assign to the Yangian generated by (3-6) a
deformation parameter equal to $ U $ {\it or} $ -U $. To fix the sign one has
to find the action of the Yangian generators upon the multiparticle eigenstates
of the hamiltonian i.e. the coproduct. Unfortunately in the present case one
apparently doesn't have means to derive the form of the symmetry generators
using the coproduct and the one-site representation of a symmetry algebra as
the starting point, as can be done, for example, in XXZ-magnet. Hence we relied
essentially upon the guesswork and heavy computations in order to get the
symmetry generators.

It is clear, that the expressions (3-6) provide a free-fermion representation
of the Yangian. How non-trivial this representation is ? One easily sees, that
taking, for example, commutators of the generators $ ( E_{1},F_{1},H_{1} ) $
beteen themselves we get operators wich contain the densities (2) with
increasing absolute values of $n$. This apparently shows that the
representation of the Yangian we are dealing with is infinite dimensional.
Moreover, when $ U=0 $ the generators (3-6) turn into the linear combinations
of the infinite-wedge (infinite-dimensional)[13] representation for the
generators of $ sl(2) $-Kac-Moody algebra.

Is the above Yangian symmetry complete ? Probably it's not, and can be extended
to the Yangian Double which is a deformation of the complete $ sl(2) $-Kac
Moody algebra [10]. Indeed in the case, for example, of XXZ magnet the complete
algebra is a deformation of the complete ${\hat sl(2)} $ so it is natural to
look for the larger symmetry in the case of the Hubbard model as well.

\section{Conclusion}
As we have seen the 1-d nearest-neighbour interaction  Hubbard model has the
Yangian plus Yangian symmetry. This symmetry is an extension of $ sl(2) \oplus
sl(2) $. The latter symmetry is generic to the Hubbard model and exists not
only in the simplest case we were looking into, but also in higher-dimensional
cases and in the Hubbard models on non-periodic lattices [12]. It would be
interesting to see if these more comlicated Hubbard models admit some
extensions of their $ sl(2) \oplus sl(2) $ symmetries.

Returning to the model (1) the next  step is to try to utilise the Yangian
symmetry in order to get the correlators and the form-factors. This includes
the identification of the Hilbert space of the model as a representation of the
Yangians and expression of the elementary excitations' creation opertors in
terms of vertex operatos related to the Yangians.

It will be instructive to understand the relation between the Yangian symmetry
and the abelian algebra of conservation laws discovered in [14].

It also would be interesting to try to modify the Hubbard model ( by the
introduction of some non-local interaction ) in such a way that a new model
would exhibit a Yangian symmetry on the finite lattice. This program was
recently performed for the Heisenberg model in [15].

\vspace{0.5cm}

\begin{large}
{\bf Acknowlegements}
\end{large}
D.B.U. is grateful to Prof. M. Rocek for support. This work was supported in
part by the NSF grant 9309888 and by NATO 9.15.02 RG 901098 Special Panel on
Chaos, Order and Patterns: ``Functional Integral Methods in Statistical
Mechanics and Correlations''.
\vspace{1cm}

\begin{large}
{\bf References } \end{large} \\
1. E.H. Lieb and F.Y. Wu, Phys. Rev. {\bf 20}, 1445 (1968).\\
2. C.N. Yang, Phys. Rev. Lett. {\bf 19}, 1312 (1967).\\
3. E.H. Lieb, E.H., in Proceedings of the conference ``Advances in Dynamical
Systems and Quantum Physics'', Capri, May 1993 (to be published by World
Scientific); V.E. Korepin and F.H. Essler eds., {\em Exactly Solvable Models of
Strongly Correlated Electrons} (to be published by World Scientific); A.
Montorsi ed., {\em The Hubbard Model} ( World Scientific, Singapore, New
Jersey, London, Hong Kong, 1992); Baeriswyl, Campbell, Carmelo, Guinea and
Louis eds., Proceedings of the conference ``Physics and Mathematical Physics of
the Hubbard model'', San Sebastian, Spain, 1993 (to be published by Plenum). \\
4. F.H.L. Essler and V.E. Korepin, Institute for Theoretical Physics, Stony
Brook, Preprint ITP-SB-93-40, 1993 (to be published).\\
5. H.J. Schultz, Int. J. Mod. Phys. B {\bf 5}, 51 (1991); H. Frahm and V.
Korepin, Phys. Rev. B {\bf 43}, 5653 (1991). \\
6. B. Davies, O. Foda, M. Jimbo, T. Miwa and A. Nakayashiki, Comm. Math. Phys.
{\bf 151}, 89 (1993). \\
7. M. Jimbo, K.  Miki, T. Miwa and A. Nakayashiki, Phys. Lett. A {\bf 168}, 256
(1993).\\
8. M. Idzumi, T. Tokihiro, K. Iohara, M. Jimbo, T. Miwa  and T. Nakashima, Int.
J. Mod. Phys. A {\bf 8}, 1479 (1993).\\
9. V.G. Drinfel'd, in {\em Proceedings of ICM} ( Berkeley, CA , 1986). \\
10. A. LeClair and F.A. Smirnov, Int. J. Mod. Phys. A {\bf 7}, 2997 (1992).\\
11. V. Chari and A. Pressley, L'Enseignement Math\'{e}matique {\bf 36}, 267
(1990).\\
12. E.H. Lieb and O.J. Heilmann, Ann. New York Acad.Sci. {\bf 172}, 583 (1971);
C.N. Yang, Phys. Rev. Lett. {\bf 63}, 2144 (1989); Phys. Lett. A {\bf 161}, 292
(1991); M. Pernici, Europhys. Lett. {\bf 12}, 75 (1990).  \\
13. V.G. Kac, {\em Infinite dimensional Lie Algebras }( Cambridge University
Press, Cambridge, 1990).\\
14. B.S. Shastry, Journ. Stat. Phys. {\bf 50}, 57 (1988). \\
15. D. Bernard, M. Gaudin, F.D.M. Haldane, V. Pasquier, `` Yang-Baxter Equation
in Spin Chains with Long Range Interaction `` Preprint SPhT-93-006.
\end{document}